\documentclass[12pt]{article}
\usepackage{amssymb}
\columnwidth\textwidth

\begin{document}

\newcommand{\be}{\begin{equation}}
\newcommand{\ee}{\end{equation}}
\newcommand{\beann}{\begin{eqnarray*}}
\newcommand{\eeann}{\end{eqnarray*}}
\newcommand{\bea}{\begin{eqnarray}}
\newcommand{\eea}{\end{eqnarray}}
\newcommand{\lb}{\label}
\begin{titlepage}

\noindent

\vspace*{1cm}
\begin{center}
{\large\bf HAWKING RADIATION FROM DECOHERENCE} 

\vskip 1cm

{\bf Claus Kiefer} 
\vskip 0.4cm
Institut f\"ur Theoretische Physik,\\ Universit\"{a}t zu K\"oln, \\
Z\"ulpicher Str.~77,
50937 K\"oln, Germany.\\
\vspace{1cm}

\nopagebreak[4]

\begin{abstract}
It is argued that the thermal nature of Hawking radiation arises
solely due to decoherence. Thereby any information-loss paradox
is avoided because for closed systems pure states remain pure. 
The discussion is performed for a massless scalar field in
the background of a Schwarzschild black hole, but the arguments
should hold in general. The result is also compared to and contrasted
with the situation in inflationary cosmology.
\end{abstract}
\end{center}

\end{titlepage}

The process of black-hole evaporation is still not understood
on the most fundamental level. One of the key questions concerns
the ``information-loss paradox'' -- can a pure quantum state 
evolve into a mixed state during the evaporation or not \cite{Page}?
In this Letter I shall argue that Hawking radiation always remains in
a pure state and that its mixed appearance emerges through the
irreversible process of decoherence \cite{deco}.

Consider the simplest case of a massless scalar field $\phi$.
In the situation of an Unruh observer in Minkowski spacetime one 
usually considers a hypersurface of constant Rindler time
which connects the left with the right Rindler wedge. Tracing out
in the Minkowski vacuum the modes of the left part leads to a
density matrix in the right part corresponding to a canonical
ensemble with temperature $a/2\pi$, where $a$ is
the acceleration \cite{Is}. Alternatively, one can impose the
boundary condition $\phi=0$ at the origin (corresponding to the
presence of a ``mirror''). 
In this case the evolution along the right part of constant Rindler
time (called $t$) hypersurfaces is given by the quantum state
(denoting the Fourier transform of the scalar field by $\phi(k)$)
\cite{KHM}
\be
\Psi  \propto \exp\left[-\int_{-\infty}^{\infty}
      dk\ k\ {\rm coth}\left(\frac{\pi k}{2a}+ikt\right)
      \vert\phi(k)\vert^2\right] \ . \lb{unruh}
\ee
(The usual normalisation factor for a Gaussian is being assumed.)
The corresponding collapse-situation in the black-hole case
was considered in \cite{DK} within the CGHS model of two-dimensional
gravity. The same quantum state as (\ref{unruh}) was found outside
the hole, with $a$ being replaced by the CGHS parameter $\lambda$.
It is also the case for the eternal hole with the boundary condition
$\phi=0$ at the bifurcation sphere.
I shall assume that the same result holds for a four-dimensional
Schwarzschild black hole with $a$ being replaced by $\kappa
\equiv(4M)^{-1}$, where $M$ denotes the mass of the black hole
and $t$ is the Schwarzschild time.
The expectation value of the number operator for the mode ${\mathbf k}$
in this state is then given by the Planck form, independent of $t$,
\be
\langle n_{\mathbf k}\rangle = \frac{1}{e^{8\pi\omega M}-1}\ ,
\lb{number}
\ee
where $\omega=\vert{\mathbf k}\vert\equiv k$. 

It was shown in \cite{GS} that the vacuum quantum state in a black-hole
spacetime is given, for each mode, by a two-mode squeezed  
vacuum. Such a state also results from an inflationary phase of the
early universe. The squeezing parameter is given by
\be
\tanh r_{k}=\exp(-4\pi\omega M)\ . \lb{r}
\ee
(Note that the imaginary part of the action for an s-wave outgoing
particle in the Hawking radiation is given by $4\pi\omega
(M-\omega/2)$, being equal to $-1/2$ times the change of the Bekenstein-Hawking
entropy \cite{PW}.) 
One recognises that $r_k\to 0$ ($r_k\to\infty$) for $k\to\infty$
($k\to 0$), i.e. there is no squeezing for small wavelengths, but high
squeezing for large wavelengths. For the frequency at the maximum of
the Planck spectrum one finds from Wien's law that the
squeezing parameter is $r_k\approx 0.25$ corresponding to
an expectation value of particle number $\langle n_k\rangle
=\sinh^2r_k\approx 0.06$. This is drastically different from the
situation in inflationary cosmology where the distribution is
not of Planck form and where squeezing
 becomes very high for the relevant modes.
I assume in the following that space is finite
(the black hole being inside a large box), so that one can write
$\Psi=\prod_{\mathbf k}\psi_{\mathbf k}$ with
\be
\psi_{\mathbf k}\propto \exp\left[-k\ {\rm coth}(2\pi kM+ikt)
 \vert\phi(k)\vert^2\right]\ . \lb{psik}
\ee
That this does in fact correspond to a squeezed state can be
recognised by writing (\ref{psik}) in the form
\be
\psi_{\mathbf k} \propto\exp\left[-k\frac{1+e^{2i\varphi_k}\tanh r_k}
 {1-e^{2i\varphi_k}\tanh r_k}\vert\phi(k)\vert^2\right]
\equiv \exp\left[-(\Omega_R+i\Omega_I)\vert\phi(k)\vert^2\right]\ ,
\lb{squeezed}
\ee
with the squeezing parameter $\tanh r_k=\exp(-4\pi\omega M)$
according to (\ref{r}) and the squeezing angle $\varphi_k=-kt$.

A squeezed state can also be characterised by the contour of the
corresponding Wigner function in phase space, which exhibits
explicitly both direction and amount of squeezing. 
For a state of the form (\ref{squeezed}), the Wigner function 
for each mode reads (cf. \cite{PS,KLPS,KPS} for the analogous situation in
cosmology)
\be
W(\phi,p)=\frac{1}{\pi}\exp\left[-\frac{2(p+\Omega_I)^2}{\Omega_R}
 -2\Omega_R\phi^2\right]\ , \lb{wigner}
\ee
where $\phi$ and $p$ denote one mode of the field and its momentum,
respectively (out of the two real modes in the complex field).
The momentum is peaked around its classical value
$p_{cl}=-\Omega_I$ with width $\Omega_R^{1/2}$, while the field
mode itself is peaked around zero with width $\Omega_R^{-1/2}$.
Denoting $4\pi\omega M\equiv x>0$ one has in particular
\be
\Omega_R(t)=\frac{k(1-e^{-2x})}{1+e^{-2x}-2e^{-x}\cos(2kt)}\ .
\lb{OmegaR}
\ee
Evaluating this expression at $t=0$ one finds $\Omega_R(0)>k$,
so that, compared to the ground state
where $\Omega_R=k$, the state is squeezed in $\phi$.
Evaluation at $kt=\pi/2$ leads to $\Omega_R(\pi/2k)<k$, so the squeezing
is in the $p$-direction.
 One has $\Omega_R(\pi/2k)/\Omega_R(0)=\tanh^2(2\pi kM)$,
which for the frequency corresponding to the maximum of the
Planck spectrum gives $\approx 0.37$.
The Wigner ellipse rotates around the origin, and the typical timescale
of the exchange of squeezing between $\phi$ and $p$ is given 
(again for the frequency of the maximum) by
\be
t_k=\frac{\pi}{2k}\approx 14M\ , \lb{time}
\ee
which is much smaller than typical observation times. It is for this 
reason that a coarse-graining with respect to the squeezing angle
can be performed. Squeezed states are extremely sensitive to interactions
with environmental degrees of freedom \cite{deco}.
 In the present case of quickly
rotating squeezing angle this interaction leads to a diagonalisation
of the reduced density matrix with respect to the particle-number basis
\cite{Pro}. Thereby the local entropy is maximised, corresponding
to the coarse-graining of the Wigner ellipse into a circle.
The value of this entropy is given by 
\be
S_k=(1+n_k)\ln(1+n_k)-n_k\ln n_k \stackrel{r_k\gg 1}{\longrightarrow}
2r_k \ . \lb{entropy}
\ee
The integration over all modes gives $S=(2\pi^2/45)T_{BH}^3V$, which is
just the entropy of the Hawking radiation with temperature
$T_{BH}=(8\pi M)^{-1}$. In this way, the pure squeezed state
becomes indistinguishable from a canonical ensemble with temperature
$T_{BH}$. The situation is very different in inflationary cosmology
where the rotation of the Wigner ellipse is slow, corresponding
to the age of the universe for the largest
cosmological scales \cite{KLPS}. The entropy is
there much smaller than its maximal value $2r_k$, which manifests
itself in the presence of acoustic peaks in the anisotropy spectrum
of the cosmic microwave background \cite{KPS}.

Independent of this practical indistinguishability
from a thermal ensemble, the state remains
a pure state. In fact, for timescales smaller than $t_k$ the
above coarse-graining is not allowed and the difference to a thermal state
could be seen in principle. In the case of a primordial black hole
with mass $M\approx 5\times 10^{14}\ {\rm g}$, the time is
$t_k\approx 1.7\times 10^{-23}\ {\rm s}$, which could be of observational
significance. 

The above consideration refers to hypersurfaces of constant $t$ which
remain outside the horizon. Observations that are performed far outside the
horizon should, however, not depend on the location of 
spacelike hypersurfaces close to the black hole. The above
arguments should thus also hold for hypersurfaces which enter the
black horizon. One can mimic this situation by considering
for simplicity a hypersurface of constant $t$ in
an eternal hole. The Minkowski vacuum $\Psi_M$ along such a surface
connecting regions $III$ and $I$ in the Kruskal diagram
can be written for $t=0$
 as $\Psi_M=\prod_{\mathbf k}\psi_{\mathbf k}$ with \cite{KHM}
\bea
\psi_{\mathbf k}\propto & &  \exp\left[-k\ {\rm coth}\left(4\pi kM\right)
 \left(\vert\phi_{III}(k)\vert^2 +\vert\phi_I(k)\vert^2\right)
 \right.\nonumber\\
& & \ \left. -\frac{k}{\sinh(4\pi kM)}(\phi^*_{III}(k)\phi_I(k)
 +\phi^*_{I}(k)\phi_{III}(k))\right]\ , \lb{full} 
\eea
where $\phi_I$ and $\phi_{III}$ denote the modes of the scalar field
in the Kruskal regions $I$ and $III$, respectively. (Note the occurrence
of $4\pi kM$ instead of $2\pi kM$ in (\ref{psik}).) As is well known,
integrating out the modes $\phi_{III}$ from (\ref{full}) leads to a
thermal density matrix in $I$ with temperature $T_{BH}$
\cite{Is}. This is of course
due to the fact that the Minkowski vacuum is an entangled state correlating
regions $I$ and $III$. It is now easy to show that the state
(\ref{full}) can be written as a product of two squeezed states.
Making a unitary transformation to a new basis,
\be
\left(\begin{array}{c}\phi_{I}\\ \phi_{III} \end{array} \right)
=\frac{1}{\sqrt{2}}\left(\begin{array}{c}\chi_1-\chi_2\\ \chi_1+\chi_2
\end{array} \right)\ , \lb{rotation}
\ee
(\ref{full}) becomes
\be    
\psi_{\mathbf k}\propto \exp\left[-k\ {\rm coth}(2\pi kM)\vert\chi_1(k)\vert^2
-k\ {\rm tanh}(2\pi kM)\vert\chi_2(k)\vert^2\right]
 \equiv \psi_1[\chi_1]\otimes\psi_2[\chi_2]\ . \lb{prod}
\ee
The state $\psi_1$ directly corresponds to the squeezed state 
(\ref{psik}), while $\psi_2$ follows from this state by the
replacement $\varphi_k\to\varphi_k+\pi/2$ for the squeezing angle.
The arguments above concerning decoherence remain true.

Since a mixed state for a closed system
in this line of arguments never occurs,
 one can expect that unitarity is preserved
during the whole black-hole evolution. The above formalism
is of course only valid as long as the gravitational background
remains fixed, but it is not expected that the inclusion of back reaction
changes this scenario. There is thus no ``information-loss paradox''
in the first place.  

The above discussion has not yet addressed the issue of the
Bekenstein-Hawking entropy $S_{BH}$. One might think, however, that
$S_{BH}$ could also be understood along these lines. For this purpose
one would need the inclusion of the correct {\em gravitational} wave function,
at least within the semiclassical approximation
(see e.g. \cite{BH} for some attempts in this
direction). The universal nature
of $S_{BH}$ would then arise due to the decohering influence
of environmental degrees of freedom on this wave function.
It has even been argued that black holes come into existence
only by decoherence \cite{My}. The universality should then hold independent
of the existing microscopic degrees of freedom, such as D-branes, 
at least for black holes that are much heavier than the Planck mass.

\end{document}